\documentclass[11pt]{article}
\pdfoutput=1

\usepackage{aas_macros,amsmath,amssymb,comment,cite,esint,graphicx,mathtools}
\usepackage[margin=.8in,letterpaper]{geometry}
\usepackage[colorlinks=true]{hyperref}
\usepackage[affil-it]{authblk}
\usepackage{url}

\numberwithin{equation}{section}
\let\originalleft\left
\let\originalright\right
\renewcommand{\left}{\mathopen{}\mathclose\bgroup\originalleft}
\renewcommand{\right}{\aftergroup\egroup\originalright}
\mathcode`\*="8000
{\catcode`\*=\active\gdef*{\mathclose{}\,\mathopen{}}}

\newcommand{\ab}[1]{\left|#1\right|}
\newcommand{\br}[1]{\left[#1\right]}

\newcommand{\pa}[1]{\left(#1\right)}
\newcommand{\td}[1]{\tilde{#1}}

\newcommand{\be}{\begin{equation}}
\newcommand{\ee}{\end{equation}}
\newcommand{\bea}{\setlength\arraycolsep{2pt} \begin{eqnarray}}
\newcommand{\eea}{\end{eqnarray}}
\newcommand{\nn}{\nonumber}

\def \ve {\varepsilon}

\def \k {\kappa}

\def \f {\frac}

\def \nn {\nonumber}

\begin{document}
\title{Observability of Zero-angular-momentum Sources Near Kerr Black Holes}

\author{
Haopeng Yan$^{1}$, Minyong Guo$^{1\ast}$,
Bin Chen$^{1,2,3}$}
\date{}

\maketitle

\vspace{-10mm}

\begin{center}
{\it
$^1$Center for High Energy Physics, Peking University,
No.5 Yiheyuan Rd, Beijing 100871, P. R. China\\\vspace{4mm}

$^2$Department of Physics, Peking University, No.5 Yiheyuan Rd, Beijing
100871, P.R. China\\\vspace{4mm}

$^3$ Collaborative Innovation Center of Quantum Matter,
No.5 Yiheyuan Rd, Beijing 100871, P. R. China\\\vspace{2mm}
}
\end{center}

\vspace{8mm}

\begin{abstract}
We revisit monochromatic and isotropic photon emissions from the zero-angular\linebreak-momentum sources (ZAMSs) near a Kerr black hole.
 We investigate the  escape probability of the photons that can reach to infinity and study the energy shifts of these escaping photons, which could be expressed as the functions of the source radius and the black hole spin. We study the cases for generic source radius and black hole spin, but we pay special attention to the near-horizon (near-)extremal Kerr ((near-)NHEK) cases. We reproduce the relevant numerical results using a more efficient method and get new analytical results for (near-)extremal cases.
The main non-trivial results are: in the NHEK region of a (near-)extremal Kerr black hole, the escape probability for a ZAMS tends to $\frac{7}{24}\approx 29.17\%$,  independent of the NHEK radius; at the innermost of the photon shell (IPS) in the near-NHEK region, the escape probability for  a ZAMS tends to
    \be
    \frac{5}{12} -\frac{1}{\sqrt{7}} + \frac{2}{\sqrt{7}\pi}\arctan\frac{1}{\sqrt{7}}\approx12.57\% .
    \nn
    \ee
The results show that the photon escape probability remains a relatively large nonzero value even though the ZAMS is in the deepest region of a near-horizon throat of a high spin Kerr black hole, as long as the ZAMS is outside the IPS.
The energies of the escaping photons at infinity, however, are all redshifted but still visible in principle.

\end{abstract}

\vfill{\footnotesize Email:haopeng.yan@pku.edu.cn,\,minyongguo@pku.edu.cn,\,bchen01@pku.edu.cn.\\$~~~~~~\ast$ Corresponding author.}

\maketitle

\baselineskip 18pt
\section{Introduction}\label{sec:intro}
The first image of M87* photographed by the Event Horizon Telescope (EHT) collaboration \cite{Akiyama:2019cqa} strongly reveals that there might be a supermassive black hole in the center of the M87 galaxy. This opens up a remarkable new avenue towards probing strong gravity region with electromagnetic observations.
A key feature of the black hole image is a dark shadow surrounded by a bright ring, which corresponds to the photon shell  consisting of spherical photon orbits. Thus understanding the image structure relies on the study of photon emissions in the exterior of the black hole, especially in the vicinity of the black hole.
Moreover,
because of the strong gravitation of a black hole, certain amount of the photons emitted from an exterior source will be captured by the black hole so that the source is only partially visible to distant observers. Once a source crosses the event horizon, no photons emitted from it can escape and thus the source can never be seen by distant observers.
The photon emissions are closely related to the astrophysical details of the near-horizon sources. In particular,  the observability of these sources, depending on the  probabilities of the photons escaping from the black hole and the energy shifts of these escaping photons, is affected by their motions. 

Many black holes are thought to be illuminated by accretion disks surrounding them. A toy model of thin accretion disk are composed of photon emitters that orbit around a black hole on equatorial circular geodesics which terminate at the innermost stable circular orbit (ISCO). However, under slight perturbations an ISCO will transit into a plunging geodesic which falls toward to the black hole center.
Besides that, other free near-horizon light sources follow the geodesics described by their integrals of motions. On the other hand, studies on non-geodesic sources are also important to understand the physics of astronomical black holes.  The locally nonrotating frame (LNRF) \cite{Bardeen:1972fi} are usually introduced to explore physical processes near a rotating black hole, since the rest sources in the LNRF, which we call the zero-angular-momentum sources (ZAMSs), are defined to corotate with the spacetime so that the ``frame-dragging" effect of the black hole can be canceled out as much as possible.  For example, the famous collisional Penrose processes are well measured in the ZAMSs frame \cite{Penrose:1969pc,Piran1975,Banados:2009pr, Guo:2016vbt, Berti:2014lva,Schnittman:2014zsa,Zhang:2020tfz}. Therefore, despite the ZAMS is introduced as an idealized toy model and not along geodesics, it can nevertheless serve as a nice reference for studying the observability of near-horizon sources with various kinds of geodesic motions.  For example, it seems that \cite{Igata:2021njn}, comparing to the photon emission from a ZAMS, a nonzero angular velocity of a source will increase the photon escaping probability while an inward radial velocity of a source will suppress the probability. As a result, the study of photon escaping probability emitted from ZAMSs is of  great value. In recent years, these kinds of near-horizon sources have attracted increasing attentions and have been studied from many observational aspects \cite{Porfyriadis:2014fja,Compere:2017hsi,Castro:2021csm
,Lupsasca:2014pfa,Compere:2016xwa,Porfyriadis:2016gwb,
Gralla:2017ufe,Lupsasca:2017exc,Ogasawara:2019mir,Igata:2019hkz,Igata:2021njn,
Gates:2020els,Gates:2020sdh,Zhang:2020pay,Cardoso:2021sip}.

Photon emissions from ZAMSs near a Kerr(-Newmann) black hole were first studied in \cite{Ogasawara:2019mir} last year, where the photon motions were exhaustively classified and the photon escape probabilities were produced numerically.
The authors found that, as a ZAMS approaches the horizon, the photon escape probability tends to approximately $29\%$ for an extremal Kerr black hole, while it keeps a relatively large nonzero value before getting to zero at the horizon for a near-extremal Kerr black hole. Soon after that,
the photon emissions from the ISCO of a Kerr black hole \cite{Igata:2019hkz} and the ones from inside the ISCO were studied\cite{Igata:2021njn}.\footnote{Recently, the light ring and the appearance of a bright light source falling into Schwarzschild black holes was also studied in Ref.~\cite{Cardoso:2021sip}.} In addition, the photon emissions from circular orbiters had also been reinvestigated in \cite{Gates:2020els,Gates:2020sdh} by introducing the ``critical curve" in the orbiter's sky which distinguishes the photon-escaping region  from  the photon-captured region. In \cite{Gates:2020els}, special attention had been paid to the high-spin case and a high-spin perturbation expansion method was applied  to analytically compute the escape probability and the redshift-dependent total fluxes of the photon emissions from the ISCO.
Main conclusions for the photon emissions from the circular sources are that \cite{Igata:2019hkz,Gates:2020els}, the escape probabilities are always greater than $50\%$ regardless of the black hole spin and orbital radius, and there are always sufficient energies carried by these escaping photons due to the fact that the Doppler boost can overcome the gravitational redshift in that case. Moreover, the escape probability has a minimum at the ISCO of an extremal black hole, which is still approximately $55\%$.

\begin{figure}[t]
  \centering
  \includegraphics[width=16.5cm]{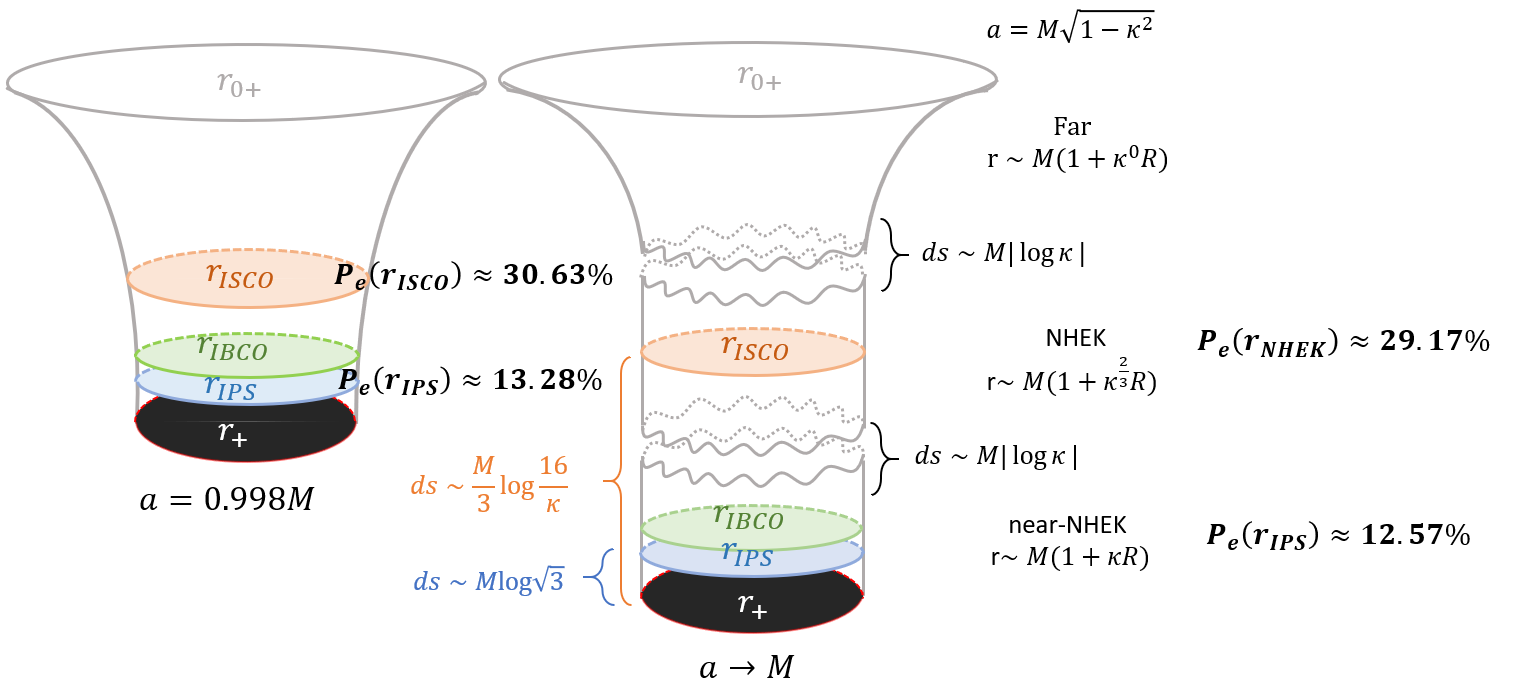}
  \caption{Throat geometry of a (near-)extremal Kerr black hole with $a=M\sqrt{1-\kappa^2}$ and $0<\kappa\ll1$ (see Appendix~\ref{app:nhgeometry} for an introduction) \cite{Bardeen:1972fi}. As $a\rightarrow M$, the near horizon region develops a infinitely deep throat with the entrance at the ergosurface $r_{0+}=2M$. The other radii $r_+$, $r_{\text{IPS}}$, $r_{\text{IBCO}}$ and $r_{\text{ISCO}}$ are, respectively, for the event horizon, the innermost photon shell (IPS), the innermost bound circular orbit (IBCO) and the innermost stable circular orbit (ISCO), and $dr$ denotes proper radial distances. We mark the photon escape probabilities at IPS and at ISCO (NHEK scale) with numerical results for $a=0.998M$ (the Thorne limit) and with the leading order analytical results in $\kappa$ for $a\rightarrow M$ (see Sec.~\ref{sec:ep}). We can see that the escape probabilities  all take relatively large non-zero values and tend to the minima in the extremal limit (see Table \ref{table:numericalresults}). Note that the result for $\mathcal{P}_e(r_{\text{NHEK}})$ actually takes an exact value $7/24$ for a precisely extremal black hole,  independent of a specific NHEK radius $R_{\text{NHEK}}$ due to the dilation symmetry along the throat. Note also that the proper radial distance between the horizon and the IPS is a finite value, $M\log\sqrt{3}$, to the leading order in $\kappa$, thus the IPS resides very close to the horizon and tends to a relatively deeper position in the near-horizon throat as the black hole approaches extreme $\kappa\rightarrow0$.}
  \label{fig:throat}
\end{figure}

In this paper, we revisit the observability of ZAMSs near a Kerr black hole that was first studied in \cite{Ogasawara:2019mir} by adapting the method introduced in \cite{Gates:2020els} which crucially relies on the critical curve in the source sky. While we will study the generic case for arbitrary black hole spin and source radius, like in \cite{Gates:2020els}, we will pay special attention  to the near-horizon sources of (near-)extremal black holes.
The reason is that a high-spin black hole exhibits a striking throat-like geometry in its near-horizon region, which can be described by the Near-Horizon (near-)Extremal Kerr ((near-)NHEK) metric\footnote{We use ``(near-)NHEK" to represent both the NHEK geometry described by \eqref{NHEKmetric} and near-NHEK geometry described by \eqref{nearNHEKmetric}.}. In the (near)-NHEK geometry, there is  an enhanced conformal symmetry which simplifies the near-horizon physics considerably such that analytical computations become tractable \cite{Bardeen:1999px}. On the other hand, many astrophysical black holes have been found to be rotating very fast \cite{Reynolds:2013qqa,MillerJones:2021plh}.  Therefore, not only do the high-spin black holes attract theoretical interests, but also they are astrophysically relevant. Indeed, the nice (near-)NHEK properties have been extensively applied to the analytical studies on gravitational waves \cite{Porfyriadis:2014fja,Compere:2017hsi,Castro:2021csm}, relativistic jets \cite{Lupsasca:2014pfa,Compere:2016xwa}, electromagnetic emission\footnote{Another analytic study on the photon emissions has been made in \cite{Guo:2019pte} for  rotating black hole spacetimes which is based on large D expansion.} and observational signatures \cite{Porfyriadis:2016gwb,Gralla:2017ufe,Lupsasca:2017exc,
Guo:2018kis,Yan:2019etp
,Guo:2019lur}. Note that the ISCO of a high-spin black hole is in the NHEK region, a region although appearing to approach the horizon (or ``coincides" with the horizon in the extreme case),  is however only in the sub-near-horizon region of a near-extremal black hole, and inside it there is still a near-NHEK region residing in the deepest portion of the near-horizon throat. As just reviewed above, the sources at ISCO in the NHEK region were proven to be well visible to distant observers, then what about the observability of the sources in the near-NHEK region? While a stable circular orbiter exists only on or outside the ISCO, a ZAMS could be arbitrarily close to the horizon.
Taking the advantages of the NHEK and near-NHEK geometries, we analytically compute the photon escape probabilities and the redshift factors by applying a high-spin perturbation method \cite{Gates:2020els}.
Our main results are summarized in Fig.~\ref{fig:throat}. In an extremal Kerr black hole, the escape probability for a ZAMS in the horizon limit, $r_s\rightarrow r_+$, is reproduced as an exact analytical expression, $7/24\approx29.17\%$. And we show that this value is also adapted to near-extremal cases for a ZAMS at the ISCO-scale (in the NHEK region) at the leading order of large spin expansion. For ZAMSs in the near-NHEK, we choose the radius of innermost photon shell (IPS) as a representative point, and we obtain  a relatively large nonzero escape probability  $5/12-1/\sqrt{7}+2/(\sqrt{7}\pi)\arctan(1/\sqrt{7})\approx12.57\%$, which means that considerable photons can still escape even in the deepest region of the near-horizon throat and the IPS may be treated as a specific bound for the point ``just before the horizon" as proposed in \cite{Ogasawara:2019mir}. Moreover, we discuss the redshifts of the escaping photons from ZAMSs, and we show that all the escaping photons are redshifted since the strong gravity effect of black hole dominates over the Doppler effect for a ZAMS. As illustration, we present several redshift density plots for the photons escaping from ZAMSs in Fig.~\ref{fig:RedshiftDensity}.

The remaining parts of this paper are organized as follows. In Sec.~\ref{sec:zams}, we introduce the LNRF in a Kerr geometry and introduce ZAMSs near a Kerr black hole. In Sec.~\ref{sec:photonmotion}, we study photon motions and define a source sky parameterized by local emission angles which are expressed in terms of  the impact parameters of the photons. Then we introduce critical impact parameters for unstable spherical photon orbits and define the critical angles and the critical curve of an escape cone using these  parameters.
In Sec.~\ref{sec:ep}, we study the escape probabilities of the photons emitted isotropically from ZAMSs. The escape probabilities are evaluated numerically for non-extremal black holes, and are computed analytically to the leading order in the deviation from extremality for (near-)extremal black holes. In Sec.~\ref{sec:redshift}, we study the energy shifts of monochromatic photons that escape to asymptotical infinity. We give a concise summary and several final remarks in Sec.~\ref{sec:discussion} and provide a brief review on the near horizon geometry of (near-)extremal Kerr black holes in Appendix ~\ref{app:nhgeometry}.

\section{ZAMSs near a Kerr black hole}\label{sec:zams}
The Kerr metric in the Boyer-Lindquist coordinates can be written in the form of
\be
\label{eq:Metric1}
ds^2=-e^{2\nu}dt^2+e^{2\psi}(d\phi-\omega dt)^2+e^{2\mu_1}dr^2+e^{2\mu_2}d\theta^2,
\ee
where $\nu$, $\psi$, $\mu_1$, $\mu_2$ and $\omega$ are the functions of $r$ and $\theta$, given by
\be
\label{eq:Metric2}
e^{2\nu}=\frac{\Sigma\Delta}{\Xi},\qquad
e^{2\psi}=\frac{\Xi\sin^2\theta}{\Sigma},\qquad
e^{2\mu_1}=\frac{\Sigma}{\Delta},\qquad
e^{2\mu_2}=\Sigma,\qquad
\omega=\frac{2Mar}{\Xi},
\ee
with
\be
\label{eq:Metric3}
\Delta=r^2-2Mr+a^2,\qquad
\Sigma=r^2+a^2\cos^2\theta,\qquad
\Xi=(r^2+a^2)^2-a^2\Delta\sin^2\theta.
\ee
This metric describes a rotating black hole with mass $M$ and spin $a$.
The radii for the inner and outer event horizons are obtained by solving $\Delta=0$,
\be
\label{eq:Horizon}
r_\pm=M\pm\sqrt{M^2-a^2},
\ee
and the radii for inner and outer ergosurfaces are obtained by solving $g_{tt}=0$,
\be
\label{eq:ErgoSurface}
r_{0\pm}=M\pm\sqrt{M^2-a^2\cos^2\theta}.
\ee
The region $r_+<r<r_{0+}$ is called the ergoregion.

The LNRF is associated with the zero-angular-momentum observers that co-rotate with the spacetime geometry, which is given by \cite{Bardeen:1972fi}
\be
\label{eq:LNRF}
e_{(t)}=e^{-\nu}(\partial_t+\omega\partial_\phi),\qquad
e_{(r)}=e^{-\mu_1}\partial_r,\qquad
e_{(\theta)}=e^{-\mu_2}\partial_\theta,\qquad
e_{(\phi)}=e^{-\psi}\partial_\phi.
\ee
In this paper, we will consider equatorial photon emitters that stay at rest with the LNRF and we call this kind of light sources ZAMSs.

\section{Photon motions and an escape cone in the source sky}\label{sec:photonmotion}
\subsection{Photon motions and the source sky}
A photon in the exterior of a Kerr black hole moves along a null geodesic with its four-momentum given by
\be
\label{eq:FourMomentum}
p_\nu dx^\nu=E\left(-dt+\sigma_r\frac{\sqrt{\mathcal{R}(r)}}{\Delta}dr+\sigma_\theta \sqrt{\Theta(\theta)} d\theta +\lambda d\phi \right)
\ee
where $\mathcal{R}(r)$ and $\Theta(\theta)$ are respectively the radial and angular potentials,
\bea
\label{eq:RPotential}
\mathcal{R}(r)&=&(r^2+a^2-a\lambda)^2-\Delta(\eta+(a-\lambda)^2),\\
\label{eq:ThPotential}
\Theta(\theta)&=&\eta+a^2\cos^2\theta-\lambda^2\cot^2\theta,
\eea
with $(E,\, \lambda,\, \eta)$ denoting the energy, the energy-rescaled angular momentum and the energy-rescaled Carter constant of the motion, and $\sigma_r,\, \sigma_\theta=\pm1$, representing the initial radial and polar directions of motion. The constants $\lambda$ and $\eta$ are also called the impact parameters of the photons.

\begin{figure}[h]
  \centering
  \includegraphics[width=7cm]{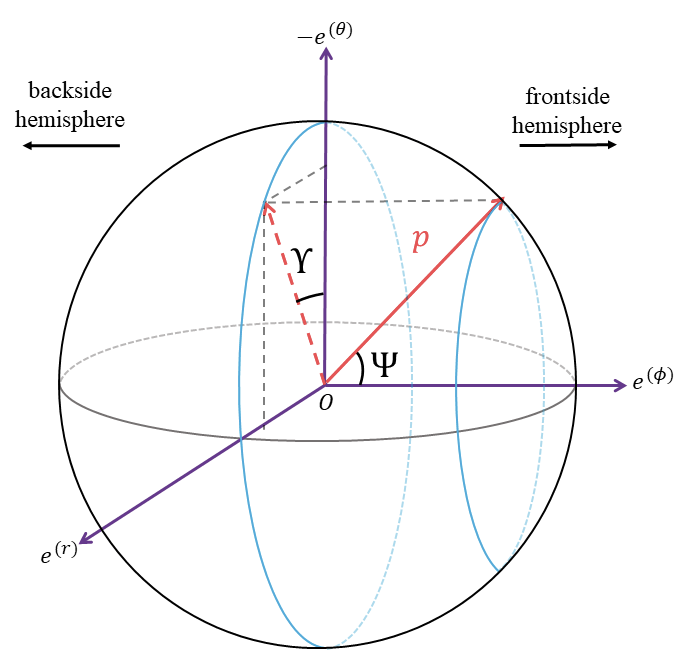}
  \caption{Source sky parameterized by local angles $\Psi$ and $\Upsilon$.}
  \label{fig:orbitersky}
\end{figure}

We now consider a ZAMS located at a radius $r_s$ on the equatorial plane $\theta_s=\pi/2$.
In the local rest frame of a ZAMS, i.e., in LNRF, we can define a pair of emission angles $(\Psi,\Upsilon)$ in the source sky by using their consines \cite{Bardeen:1973tla,Gates:2020sdh},
\bea
\label{eq:CosPsi}
\cos\Psi&=&\frac{p^{(\phi)}}{p^{(t)}}=\frac{r_s\sqrt{\Delta(r_s)}\lambda}{\xi_s-2a M\lambda},   \\
\label{eq:CosUp}
\cos\Upsilon&=&\frac{-\sigma_\theta p^{(\theta)}}{\sqrt{(p^{(r)})^2+(p^{(\theta)})^2}}
=-\sigma_\theta\frac{\sqrt{r_s \Delta(r_s)\xi_s \eta}}{r_s\sqrt{(\xi_s-2aM\lambda)^2
-r_s^2\Delta(r_s)\lambda^2}},
\eea
with $\xi_s=r_s^3+a^2(2M+r_s)$.
We treat $\Psi=\arccos[\cos\Psi]\in[0,\pi]$ as a polar angle with regard to the $+\phi$ direction,
and treat $\Upsilon=\sigma_r \arccos[\cos\Upsilon]\in(-\pi,\pi)$ as an azimuthal angle in the $r$-$\theta$ plane, a plane separating the backside ($-\phi$) and the frontside ($+\phi$) hemispheres. The emission angles in the source sky are shown in Fig.~\ref{fig:orbitersky}.

As introduced in Ref.~\cite{Gates:2020sdh}, the ``direction to the black hole center" can serve as a good reference for distinguishing the escaping photons from those captured by the central black hole. The direction corresponds to photon emissions with $\lambda=\eta=0$ and $\sigma_r=-1$. Note that it locates at
\be
\label{directiontobh}
\Psi_\bullet=\frac{\pi}{2},\qquad
\Upsilon_\bullet=-\frac{\pi}{2},
\ee
which is the same for all ZAMSs.

\subsection{Critical angles and critical curves of an escape cone}
The threshold between the photons captured by a black hole and the photons escaping to infinity corresponds to the unstable spherical photon orbits which satisfy $\mathcal{R}(\td r)=\mathcal{R}^\prime(\td r)=0$.  For these orbits,  the photon impact parameters $(\lambda,\, \eta)$ are\cite{Bardeen:1973tla}
\bea
\label{CriticalL}
\td{\lambda}(\td r)&=&a+\frac{\td r}{a}\pa{\td r-\frac{2\Delta (\td r)}{\td r-M}},\\
\label{CriticalET}
\td{\eta}(\td r)&=&\frac{\td{r}^3}{a^2}\pa{\frac{4M\Delta(\td r)}{(\td r-M)^2}-\td r}.
\eea
Since $\td{\eta}(\td r)\geq0$ for the photons crossing the equatorial plane, the orbits lie
 in the range $\td r_{\text{IPS}}\leq\td r\leq\td r_{\text{OPS}}$, where
\bea
\label{InnerShell}
\td r_{\text{IPS}}&=&2M\br{1+\cos\pa{\frac{2}{3}\arccos\br{-\frac{a}{M}}}},\\
\label{OuterShell}
\td r_{\text{OPS}}&=&2M\br{1+\cos\pa{\frac{2}{3}\arccos\br{\frac{a}{M}}}}.
\eea
The region containing all these unstable spherical photon orbits is called the photon shell, and $\td r_{\text{IPS}}$ and $\td r_{\text{OPS}}$ are the innermost photon shell (IPS) and outermost photon shell (OPS), respectively.

An escape cone consists of the solid angles of the emissions from which the photons can escape to infinity.
The boundary of an escape cone corresponds to the
critical angles in the source sky, which are obtained by plugging $(\td\lambda,\, \td\eta)$ into the local angles $[\Psi(\lambda,\eta),\Upsilon(\lambda,\eta)]$. Hereafter, the quantities adorned with a tilde are evaluated at the critical parameters \eqref{CriticalL} and \eqref{CriticalET}.

\begin{figure}[t]
  \centering
  \includegraphics[width=17cm]{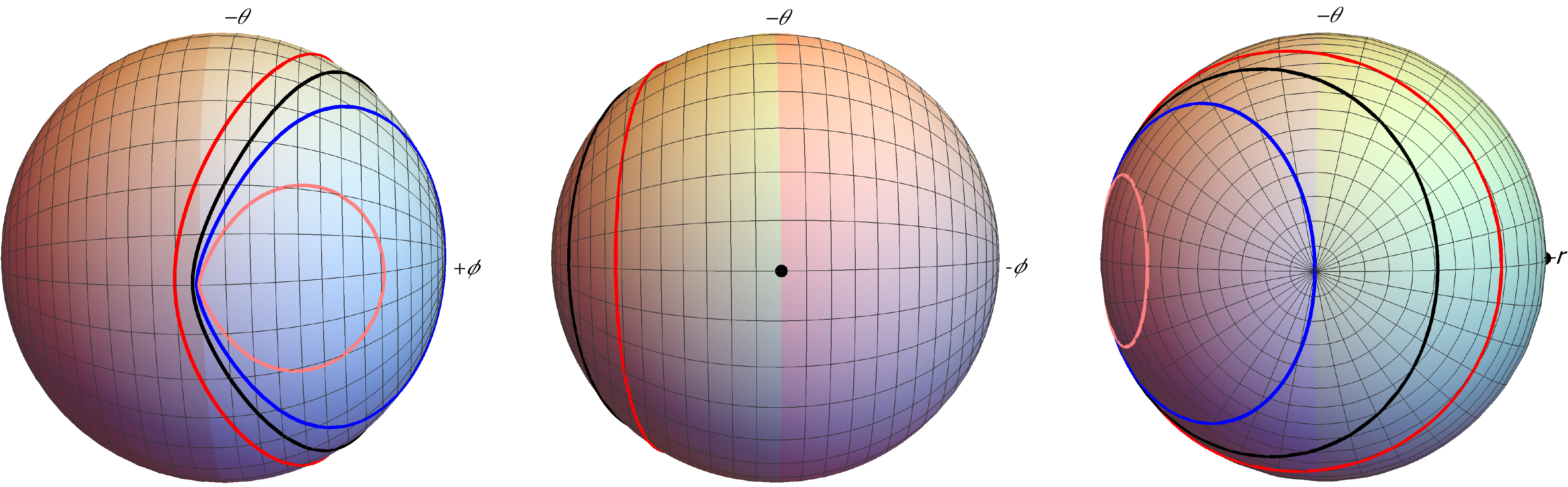}
  \caption{Escape cones for the photons emitted from ZAMSs near a Kerr black hole with $a=0.998M$ [the Thorne limit, $\kappa\approx0.063$ \eqref{SmallPara}]. Face view from the $+r$ direction (see Fig.~\ref{fig:orbitersky}): the parts of the source sky are colored by gray (left-front), blue (right-front), green (right-back) and red (left-back).  From left to right, the plots are viewed from $+r$, $-r$ and $+\phi$ directions, respectively. The orbital radii of the ZAMSs for the critical curves are $r_s=r_{\text{ISCO}}\approx1.2M$ (red), $1.1M\approx(1+2\kappa)M$ (black), $\td r_{\text{IPS}}\approx1.074M$ (blue) and $1.065M\approx (1+1.03\kappa)M$ (pink).
  The black dot in each plot represents the ``direction to the black hole center".}
  \label{fig:3dplot}
\end{figure}

We show some examples for the critical angles $(\td \Psi,\td\Upsilon)$ in the source sky with closed curves in Fig.~\ref{fig:3dplot}. We refer to these closed curves as the ``critical curves" \cite{Gates:2020sdh}.  A critical curve divides the sphere of a source sky into two parts: the one containing the ``direction to the black hole center" \eqref{directiontobh} is outside the boundary of an escape cone and is called the captured region since all photons emitted from this region are captured by the black hole, while the other complementary one is inside the boundary of an escape cone and is naturally called  the escaping region.

\subsection{Escape cone for high spin black hole}\label{sec:highspincritical}

The black hole could be of high spin. There is  an upper bound for the spin of an astrophysical black hole proposed by K. S. Thorne $a=0.998M$ \cite{Thorne:1974ve}. In this section, we study the escape cone for the (near-)extremal cases.

In the extremal limit $a\rightarrow M$, there emerges a deep throat (see Fig.~\ref{fig:throat}) near the black hole horizon to which we will pay special attention.
We provide a brief introduction to the near horizon geometry of (near-)extremal Kerr black holes in Appendix~\ref{app:nhgeometry}, and a comprehensive review may be found in Ref.~\cite{Kapec:2019hro}.

It is convenient to introduce a small parameter $0<\kappa\ll1$ to describe the deviation from extremality,
\be
\label{SmallPara}
a=M\sqrt{1-\k^2}
\ee
and to work with the dimensionless Bardeen-Horowitz \cite{Bardeen:1999px} coordinate $R$ for the radius in the throat, defined by \cite{Hadar:2014dpa,Gralla:2015rpa}
\be
\label{RsExpansion}
r=M\pa{1+\kappa^p R},\qquad 0<p\leq1.
\ee
Each specific value $p$ defines a certain ``NHEK band" which is separated with another one significantly \cite{Kapec:2019hro}. Among these NHEK bands, we have special interests in the $p=\frac{2}{3}$ band and the $p=1$ band, which are respectively the so-called NHEK band \eqref{NHEKmetric} and near-NHEK band \eqref{nearNHEKmetric}. Note that the ISCO is located in the NHEK band, while the innermost bound circular orbit (IBCO) and the event horizon are in the near-NHEK band \cite{Gralla:2015rpa,Kapec:2019hro}. The radii outside the throat are resolved by the Extreme Kerr metric
``$p=0$". For a precisely extremal Kerr black hole we have $\kappa=0$, thus all these NHEK bands coalesce into a unique NHEK which resolves the near horizon geometry while the Extreme Kerr metric fails to resolve that region since it is degenerated at the horizon (see Appendix~\ref{app:nhgeometry}).

In the extremal limit $\kappa \to 0$, the IPS \eqref{InnerShell} and OPS \eqref{OuterShell} become
\bea
\label{ExtremeInner}
\td r_{\text{IPS}}&=&M\pa{1+\kappa\td R_{\text{IPS}}+\mathcal{O}(\kappa^2)},\\
\label{ExtremeOuter}
\td r_{\text{OPS}}&=&4M\pa{1+\mathcal{O}(\kappa^2)},
\eea
where $\td R_{\text{IPS}}=2/\sqrt{3}$.
Note that the entire photon shell stretches from the near-NHEK $p=1$ region to the Extreme Kerr $p=0$ region, and the IPS sits at the same band as the event horizon (both at the near-NHEK band) with a finite separation, $ds(r_+,r_{\text{IPS}})=M\log\sqrt{3}+\mathcal{O}(\kappa)$. This means that in order to resolve the full closed critical curve in the source sky, we need to integrate over the radius of photon shell from all these scales, i.e. from near-NHEK to the Extreme Kerr. Especially, for a precisely extremal black hole, setting $a=M$ in $(\td \Psi,\td\Upsilon)$ only resolves the portion of the critical curve from the far region of Extreme Kerr, the missing portion requires to a careful study of the NHEK geometry (see Appendix~\ref{app:nhgeometry}).

Regarding to the location of the source
\be
\label{expsource}
r_s=M(1+\kappa^q R_s),
\ee
we will consider separately  the cases of the Extreme Kerr ZAMSs $(q=0)$, NHEK ZAMSs $(0<q<1)$ and near-NHEK ZAMSs $(q=1)$. As illustrated above,
in order to resolve the full escape cone, apart from the scale of source radius we need also to consider various different extremal limits for the radii of photon shell. Thus, in each scale of the ZAMS radius, we will consider various extremal limits for the radii of the photon shell \cite{Gates:2020sdh},
\be
\label{TdRExpansion}
\td r=M(1+\kappa^p \td R), \qquad 0<p\leq1.
\ee

To summarize, we will consider the high-spin expansions \eqref{SmallPara}, \eqref{expsource} and \eqref{TdRExpansion}
for the critical emission angles $(\td \Psi,\td \Upsilon)$ of a ZAMS with $q=0$ (Extreme Kerr), $q=\f{2}{3}$ (ISCO NHEK band) and $q=1$ (near-NHEK band), respectively, and for the photon shell radii with $0\leq p\leq1$. It was found in \cite{Gates:2020els} that, for the Extreme Kerr case and the NHEK case, photon emissions from the scalings $p^+=\f{q}{2}$ and $p^-=1$ complete the critical curve while those from $p^{m+}\in(0,\f{q}{2})$ and $p^{m-}\in(\f{q}{2},1)$ only contribute the points that connect the $p^\pm$ parts (see Secs.~\ref{sec:ekcritical} and \ref{sec:nkcritical}). Later we can also see from Sec.~\ref{sec:nnkcritical} that this still holds for the near-NHEK case\footnote{That is to say, one is forced to take the specific values $p^+=\f{q}{2}$ and $p^-=1$ in order to resolve the critical curve since only photon emissions from these $p^\pm$ limits contribute nontrivially to the impact parameters \eqref{CriticalL} and \eqref{CriticalET}.}. Note that for the Extreme Kerr case we have $p^+=q=0$ and $p^{m\pm}$ reduce to $p^m\in(0,1)$.
Hereafter, we use the superscripts ``$\pm$ to denote the outer/inner limits and use ``$m$" to represent the middle limit.
Next, we will give the results for the critical emission angles for each case.
In the following, we will use subscripts ``EK'', ``NK" and ``nNK" to represent the Extreme Kerr, NHEK and near-NHEK cases, respectively.

\subsubsection{Extreme Kerr ZAMSs}\label{sec:ekcritical}
For Extreme Kerr ZAMSs, we have
\be
\label{EKcondition}
q_{\text{EK}}=0,\qquad
p^+_{\text{EK}}=0,\qquad
p^-_{\text{EK}}=1,\qquad
0<p^m_{\text{EK}}<1.
\ee
in the expansions \eqref{expsource} and \eqref{TdRExpansion}.
Then, to the leading order as $\kappa\rightarrow0$, we obtain
\bea
\label{EKPsiOut}
\td{\Psi}^+_{\text{EK}}&=&\arccos\br{-\frac{R_s(R_s+1)(\td R^2-2)}{4R_s+3R_s^2+R_s^3+2\td R}},       \\
\label{EKupout}
\td{\Upsilon}^+_{\text{EK}}&=&\sigma_r\arccos\br{-\sigma_\theta\frac{R_s\sqrt{3-\td R}(1+\td R)^{3/2}}{\sqrt{(1+R_s)(3R_s^2+R_s^3+\td R^4-R_s \td R^2(\td R^2-4)}}},
\eea
and
\bea
\label{EKpsiin}
\td{\Psi}^-_{\text{EK}}&=&\arccos\br{\frac{2(1+R_s)}{4+3R_s+R_s^2}},       \\
\label{EkUpin}
\td{\Upsilon}^-_{\text{EK}}&=&\sigma_r\arccos\br{-\sigma_\theta
\sqrt{\frac{3(1-\td R_{\text{IPS}}^2/\td R^2)}
{3+4R_s+R_s^2}}},
\eea
and
\bea
\label{EKpsim}
\td{\Psi}^m_{\text{EK}}=\arccos\br{\frac{2(1+R_s)}{4+3R_s+R_s^2}},       \qquad
\td{\Upsilon}^m_{\text{EK}}=\sigma_r\arccos\br{-\sigma_\theta
\sqrt{\frac{3}
{3+4R_s+R_s^2}}},
\eea
where $R_s\in(0,\infty)$, $\td R(+)\in(0,3)$ and $\td R(-)\in[\td R_{\text{IPS}},\infty)$, and $\sigma_r=\text{sign}(\td r-r_s)$.

\subsubsection{NHEK ZAMSs}\label{sec:nkcritical}
For NHEK ZAMSs at the ISCO scale, we have
\be
\label{nkcondition}
q_{\text{NK}}=\frac{2}{3},\qquad
p^+_{\text{NK}}=\frac{1}{3},\qquad
p^-_{\text{NK}}=1,\qquad
0<p^{m+}_{\text{NK}}<\frac{1}{3},\qquad
\frac{1}{3}<p^{m-}_{\text{NK}}<1.
\ee
in the expansions \eqref{expsource} and \eqref{TdRExpansion}.
Then, to the leading order as $\kappa\rightarrow0$, we obtain
\bea
\label{nkpsiout}
\td{\Psi}^+_{\text{NK}}&=&\arccos\br{\frac{R_s}{2R_s+\td R^2}},       \\
\label{nkupout}
\td{\Upsilon}^+_{\text{NK}}&=&\sigma_r\arccos
\br{-\sigma_\theta\sqrt{\frac{3R_s^2}
{3R_s^2+4R_s\td R^2+\td R^4}}},
\eea
and
\bea
\label{nkpsiin}
\td{\Psi}^-_{\text{NK}}&=&\frac{\pi}{3},      \\
\label{nkupin}
\td{\Upsilon}^-_{\text{NK}}&=&\sigma_r\arccos\br{-\sigma_\theta\sqrt{1-\frac{\td R^2_{\text{IPS}}}{\td R^2}}},
\eea
and
\bea
\label{nkpsimp}
\td{\Psi}^{m+}_{\text{NK}}=\frac{\pi}{2}, \qquad
 \td{\Upsilon}^{m+}_{\text{NK}}=\f{\pi}{2},    \qquad
\td{\Upsilon}^{m-}_{\text{NK}}=\f{\pi}{3},  \qquad
\td{\Upsilon}^{m-}_{\text{NK}}=0,\pi.
\eea
where $R_s\in(0,\infty)$, $\td R(+)\in(0,\infty)$ and $\td R(-)\in[\td R_{\text{IPS}},\infty)$, and $\sigma_r=\text{sign}(\td r-r_s)$.

Inverting these relations, we see that the factor $\td R/R_s$ may be eliminated out and we can arrive at a unified expression for the NHEK critical curve in the source sky,
\begin{align}
\label{nkCritPsifull}
	\cos\td\Psi_{\text{NK}}=
	\begin{cases}
		 \frac{\cos\td{\Upsilon}_{\text{NK}}}{\sqrt{3+\cos^2\td{\Upsilon}_{\text{NK}}}}
\qquad& 0\leq\td\Upsilon_{\text{NK}}\leq\pi,  \\
		\frac{1}{2}
\qquad& -\pi\leq\td\Upsilon_{\text{NK}}\leq0.
	\end{cases}
\end{align}

\subsubsection{Near-NHEK ZAMSs}\label{sec:nnkcritical}
For near-NHEK ZAMSs, we have
\be
\label{nNkcondition}
q_{\text{nNK}}=1,\qquad
p^+_{\text{nNK}}=\frac{1}{2},\qquad
p^-_{\text{nNK}}=1,\qquad
0<p^{m+}_{\text{nNK}}<\frac{1}{2},\qquad
\frac{1}{2}<p^{m-}_{\text{nNK}}<1.
\ee
in the expansions \eqref{expsource} and \eqref{TdRExpansion}.
Then, to the leading order as $\kappa\rightarrow0$, we obtain
\bea
\label{nNkPsiout}
\td{\Psi}^+_{\text{nNK}}&=&\arccos\br{\frac{\sqrt{R_s^2-1}}{2R_s+\td R^2}},       \\
\label{nnkupout}
\td{\Upsilon}^+_{\text{nNK}}&=&\sigma_r\arccos\br{-\sigma_\theta
\sqrt{\frac{3(R_s^2-1)}{1+3R_s^2+4R_s \td R^2+\td R^4}}},
\eea
and
\bea
\label{nnkpsiin}
\td{\Psi}^-_{\text{nNK}}&=&\arccos\br{\frac{\td R\sqrt{R_s^2-1}}{2R_s \td R- 2}},       \\
\label{nnkupin}
\td{\Upsilon}^-_{\text{nNK}}&=&\sigma_r\arccos\br{-\sigma_\theta
\sqrt{\frac{3\td R^2(R_s^2-1)(1-\td R_{\text{IPS}}^2/\td R^2)}{4-8R_s\td R+\td R^2+3R_s^2\td R^2}}},
\eea
and
\bea
\label{nNkPsimp}
\td{\Psi}^{m+}_{\text{nNK}}=\frac{\pi}{2}, \qquad
&&\td{\Upsilon}^{m+}_{\text{nNK}}=\f{\pi}{2},\\
\td{\Psi}^{m-}_{\text{nNK}}=\arccos\br{\frac{\sqrt{R_s^2-1}}{2R_s}},\qquad 
&&\td{\Upsilon}^{m-}_{\text{nNK}}=\arccos\br{-\sigma_\theta
\sqrt{\frac{3(R_s^2-1)}{1+3R_s^2}}},
\eea
where $R_s\in(1,\infty)$, $\td R(+)\in(0,\infty)$ and $\td R(-)\in[\td R_{\text{IPS}},\infty)$, and $\sigma_r=\text{sign}(\td r-r_s)$.
Note that as a ZAMS approaches the event horizon, we have $R_s\rightarrow1$.

\begin{figure}[h]
  \centering
  \includegraphics[width=16cm]{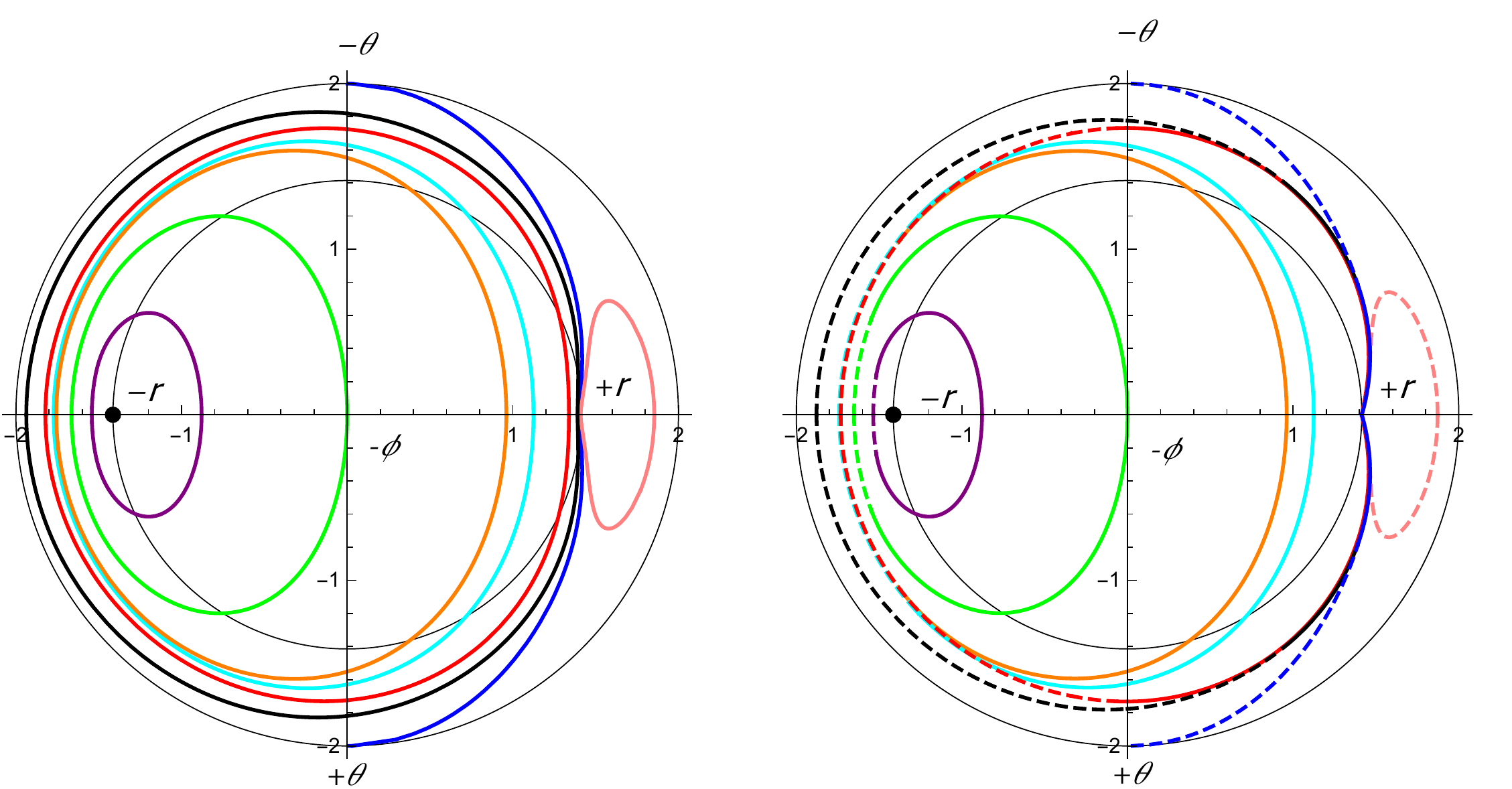}
  \caption{Escape cones of ZAMSs near high spin Kerr black holes in the backside projected plane \eqref{DefbackProjection}: the center $\rho_b=0$ has $\Psi=\pi$, the middle circle $\rho_b=\sqrt{2}$ has $\Psi=\pi/2$ and the outer circle $\rho_b=2$ has $\Psi=0$. Left: exact numerical results for a black hole with spin $a=0.998M$ [the Thorne limit, $\kappa\approx0.063$, \eqref{SmallPara}]; and the orbital radii are $10M$ (purple), $\td r_{\text{OPS}}\approx4M$ (green), $2M$ (orange, ergosphere), $1.7M$ (cyan),
  $r_{\text{ISCO}}\approx1.2M$ (red), $1.1M\approx(1+2\kappa)M$ (black), $\td r_{\text{IPS}}\approx1.074M$ (blue) and $1.065M\approx (1+1.03\kappa)M$ (pink). Right: the leading order results of high spin perturbation for $a\rightarrow M$; and the dimensionless radii $R_s$ (Eq.~\eqref{expsource}) for Extreme Kerr ZAMSs (Eq.~\eqref{EKcondition}) are $9$ (purple), $\td R_{\text{OPS}}=3$ (green), $1$ (orange, ergosphere) and $0.7$ (cyan), the radius for NHEK ZAMSs (Eq.~\eqref{nkcondition}) is $R_{\text{ISCO}}=2^{1/3}$ (red), the radii for near-NHEK (Eq.\eqref{nNkcondition}) ZAMSs are $2$ (black), $\td R_{\text{IPS}}=2/\sqrt{3}$ (blue) and $1.03$ (pink). Dashed curves are for the inner limit $p=p^-$ and solid curves are for the outer limit $p=p^+$.
   The black dot in each plot represents the ``direction to the black hole center".}
  \label{fig:EscapeCone}
\end{figure}

For a ZAMS at the IPS, $R_s=\td R_{\text{IPS}}$, we obtain a unified expression for the critical curve in the source sky,
\begin{align}
\label{nnkCritPsifull}
	\cos\td\Psi_{\text{nNK}}=
	\begin{cases}	 \frac{\cos\td{\Upsilon}_{\text{nNK}}}{\sqrt{3+\cos^2\td{\Upsilon}_{\text{nNK}}}}
\qquad& \arccos\frac{1}{\sqrt{5}}\leq\td\Upsilon_{\text{nNK}}\leq
\pi-\arccos\frac{1}{\sqrt{5}},  \\
		\frac{1+3\cos^2\td{\Upsilon}_{\text{nNK}}}{7-3\cos^2\td{\Upsilon}_{\text{nNK}}}
\qquad& 0\leq\td\Upsilon_{\text{nNK}}\leq\arccos\frac{1}{\sqrt{5}},\quad
\pi-\arccos\frac{1}{\sqrt{5}}\leq\td\Upsilon_{\text{nNK}}\leq\pi,\\
1\qquad& -\pi\leq\td\Upsilon_{\text{nNK}}\leq0.
	\end{cases}
\end{align}
Note that the third part corresponds to the point $\td\Psi=0$ in the source sky [Figs.~\ref{fig:orbitersky} and \ref{fig:3dplot}], which is the outer circle on the backside projected plane [Fig.~\ref{fig:EscapeCone}].

\section{Escape probability}\label{sec:ep}
We assume that the ZAMSs emit photons isotropically in their local rest frames. Let $\mathcal{A}_{e}/\mathcal{A}_{c}$, respectively, be the area of the escaping/captured region in the source sky with unit radius for the photons.
The probability for the photons escaping to infinity is
\be
\label{DefEP}
\mathcal{P}_e=\frac{\mathcal{A}_e}{4\pi}=\frac{1-\mathcal{A}_c}{4\pi}.
\ee
In order to compute this escape probability, we need to compute the area of $\mathcal{A}_e$, or equivalently $\mathcal{A}_c$. It is convenient to
introduce the planar polar coordinates \cite{Gates:2020sdh}
\be
\label{DefbackProjection}
\rho_{b/f}=\sqrt{2(\pm\cos\Psi+1)},\qquad
\varphi_{b/f}=\frac{\pi}{2}\mp\Upsilon,
\ee
in the backside/frontside projected planes (see Figs.~\ref{fig:EscapeCone} and \ref{fig:RedshiftDensity}) which are labeled with the subscripts $b/f$, respectively, and correspond to the upper/lower signs in the expressions, respectively. These polar coordinates are defined in a way such that
the area element $d\mathcal{A}$ on each of the planes is equal to the area element on the sphere of the source sky, $d\Omega$,
\be
\label{AreaEqCond}
d\mathcal{A}=\rho d\rho\wedge d\varphi=\sin\Psi d\Psi\wedge d\Upsilon=d\Omega.
\ee

We show some examples of critical curves associated with the critical angles  $(\td \Psi,\td \Upsilon)$ in Figs.~\ref{fig:3dplot}, \ref{fig:EscapeCone} and \ref{fig:RedshiftDensity}. Note that since the ``direction to the black hole center" \eqref{directiontobh} may be inside or outside a critical curve in the projected planes (Figs.~\ref{fig:EscapeCone} and \ref{fig:RedshiftDensity}),
the interior region of a closed critical curve is correspondingly the captured region or the escaping region, whose area $\mathcal{A}_{in}$ ($\mathcal{A}_c$ or $\mathcal{A}_e$)
can be computed by
\be
\label{AreaCapture}
\mathcal{A}_{in}=\int_{in}\rho d\rho d\phi=\int_{in}\frac{1}{2}\rho^2d\phi
=\ab{\int^{\td r_{\text{OPS}}}_{\td r_{\text{IPS}}}\td\rho^2(\td r)\frac{d\td\phi(\td r)}{d\td r}\mid_{\sigma_\theta=1}d\td r},
\ee
where we have chosen $\sigma_\theta=1$
since the escaping regions are reflection-symmetric about $\theta_s=\pi/2$, and a factor of $2$ was canceled against the factor $1/2$ in front of $\rho^2$.

\begin{figure}[h]
  \centering
  \includegraphics[width=16cm]{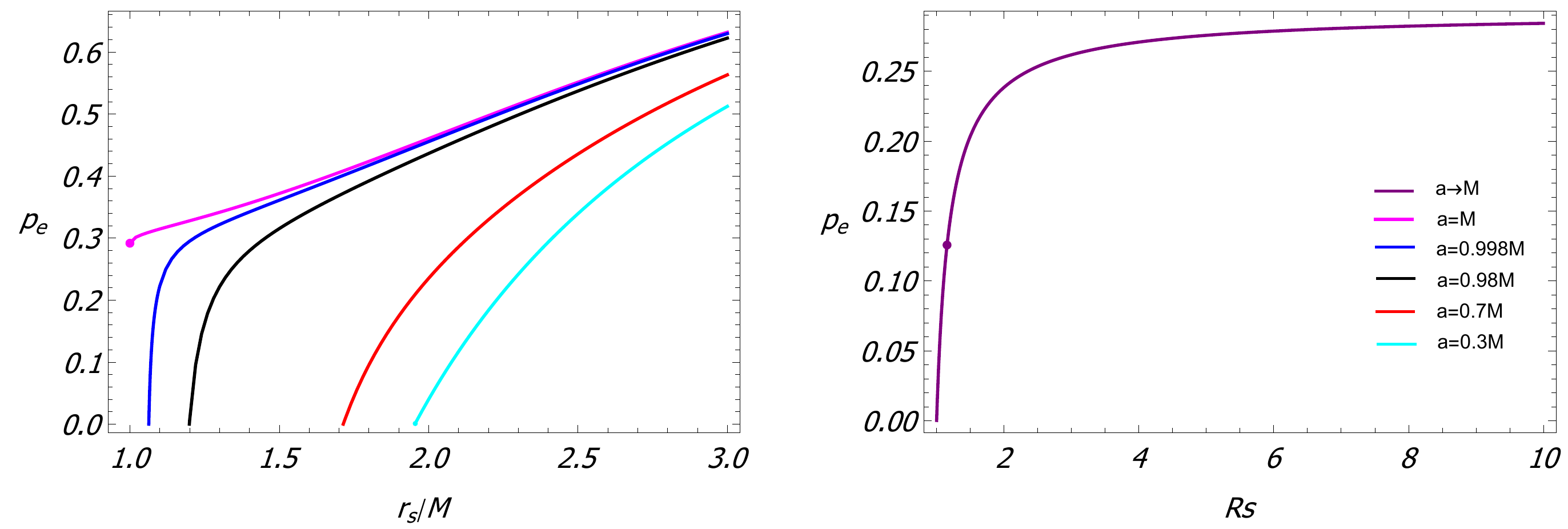}
  \caption{Left: photon escape probability from a ZAMS near a Kerr black hole of non-extremal spins [Eq.~\eqref{AreaCapture}] and extremal spin [Eq.~\eqref{areacaptureek}]. Right: leading order result in $\kappa$ for photon escape probability from a ZAMS in the near-NHEK region $r_s=M(1+\kappa R_s)$ [see Sec.~\ref{sec:nnkcritical} and Eq.~\eqref{AreaCapture}].  The magenta dot in the left figure corresponds to the critical value for a NHEK ZAMS [Eq.~\eqref{nkEP}] and the purple dot in the right figure corresponds to the critical value for a near-NHEK ZAMS at the IPS [Eq.~\eqref{nnkIMEP}]. The plots are consistent with the Figs.~6 and 8 in Ref. \cite{Ogasawara:2019mir}, but  are reproduced in a different way. Moreover, our result for a NHEK ZAMS is obtained in an exact form.}
  \label{fig:EscapeProbability}
\end{figure}

For non-extremal black holes, we deal with this integral numerically: for $r_s>r_{\text{IPS}}$ we compute $\mathcal{A}_{c}$ in the backside projected plane and for $r_s\leq r_{\text{IPS}}$ we compute $\mathcal{A}_{e}$  in the frontside projected plane. On the other hand, for (near-)extremal cases, we will compute this integral analytically to the leading order in the derivation from extremality in the next subsection. The results for the photon escape probabilities are shown in Fig.~\ref{fig:EscapeProbability}.

\subsection{Escape probabilities for high spin case}\label{sec:ephighspin}

\subsubsection{Extreme Kerr ZAMSs}\label{sec:epek}
In the backside projected area-preserving plane \eqref{DefbackProjection},  one part of the critical curve for an Extreme Kerr ZAMS (see Sec.~\ref{sec:ekcritical}) from the inner side $p=p^-$ is obtained as
\bea
\td\rho_b^-&=&\sqrt{2}\sqrt{1+\frac{2(1+R_s)}{4+3R_s+R_s^2}},\\
\td\varphi_b^-&\in&\br{-\frac{\pi}{2}+\arccos\sqrt{\frac{3}{3+4R_s+R_s^2}},
\frac{\pi}{2}-\arccos\sqrt{\frac{3}{3+4R_s+R_s^2}}},
\eea
together with the other part from the outer side $p=p^+$, the area of the escaping region \eqref{AreaCapture} can be computed by
\be
\label{areacaptureek}
\mathcal{A}_{e}=\ab{\int^{\td R_{\text{OPS}}}_{\td R_{\text{IPS}}}\td\rho_b^2(\td \Psi^+_{\text{EK}})\frac{d\td\phi_b(\td \Upsilon^+_{\text{EK}})}{d\td R}\mid_{\sigma_\theta=1}d\td R}+\br{1+\frac{(2 (1 + R_s)}{4 + 3 R_s + R_s^2}} \br{\pi-2 \arccos\frac{\sqrt{3}}{\sqrt{3 + 4R_s + R_s^2}}}.
\ee
Then plugging this into Eq.~\eqref{DefEP} gives the escape probability for an Extreme Kerr ZAMS. The result is plotted as a function of the ZAMS radius in Fig.~\ref{fig:EscapeProbability} (magenta curve).

\subsubsection{NHEK ZAMSs}\label{sec:epnk}
In the backside projected area-preserving plane \eqref{DefbackProjection}, the critical curve for a NHEK ZAMS [Eq.~\eqref{nkCritPsifull}] is obtained as
\begin{align}
\label{nkCritRho}
	\td\rho_b=
	\begin{cases}
	\sqrt{2+\frac{2\sin\td\varphi_b}{\sqrt{3+\sin^2\td\varphi_b}}}\qquad
&-\frac{\pi}{2}\leq\td\varphi_b\leq\frac{\pi}{2},	 \\
		\sqrt{3}\qquad
&\frac{\pi}{2}\leq\td\varphi_b\leq\frac{3\pi}{2}  .
	\end{cases}
\end{align}
Then from Eq.~\eqref{DefEP} and Eq.~\eqref{AreaCapture} we obtain
\be
\label{nkEP}
\mathcal{P}_e(r_{\text{NHEK}})=\frac{7}{24}\approx29.17\%.
\ee
This result is exact for a precisely extremal black hole which agrees with the numerical result first produced in Ref.~\cite{Ogasawara:2019mir}. On the other hand, the result can also be treated as the leading order result in $\kappa$ \eqref{SmallPara} for a near-extremal black hole. Note that this escape probability is independent of the NHEK radius due to the dilation symmetry of the NHEK geometry (see Fig.~\ref{fig:throat} and Appendix~\ref{app:NHEK}). For a precisely extremal black hole, the NHEK region is infinitely deep and thus the escape probability remains non-zero no matter how close to the event horizon of the ZAMS could be. For a near-extremal black hole, this means that as the ZAMS approaches the horizon, the escape probability remains greater than $29\%$ until it cross the ISCO (it tends to $29.17\%$ when the ZAMS is at the ISCO).
We plot this fixed point in Fig.~\ref{fig:throat} and Fig.~\ref{fig:EscapeProbability} (magenta dot) and show a number of numerical results for various near-extremal spins in Table \ref{table:numericalresults} (we show also an exception of high spin, 0.7, for comparison).

However, even though the ISCO of a near-extremal black hole resides in the near horizon region, it is still not very close to the horizon. In fact, the proper radial distance between the ISCO and the horizon is divergent as $\log\kappa^{-1}$ [see Fig.~\ref{fig:throat} and Eq.~\eqref{rhrisco}] \cite{Bardeen:1972fi}. This motivates us to consider the ZAMSs in the very near horizon region of a high spin black hole, i.e., in the near-NHEK region, and to see the behavior of the escape probability in that region.

\begin{table}[t]
\centering
\begin{tabular}{c c c c c c c c}
  \hline  \hline
  $a/M$&  $\rightarrow$1 & 0.9999& 0.998&0.995&0.98&0.95&0.7 \\
  \hline
  $\mathcal{P}_{\text{ISCO}}\, (\%)$ &  29.17&29.68&30.63&31.53&34.72&39.58&63.67\\
    \hline
   $\mathcal{P}_{\text{IPS}}\, (\%)$& 12.57&12.82&13.28&13.59&14.47&15.67&24.23\\
  \hline  \hline
\end{tabular}
  \caption{The first column gives the leading order analytical value in $\kappa$ \eqref{SmallPara} of escape probabilities [Eq.~\eqref{areacaptureek} and Eq.~\eqref{nkEP}] for near-extremal black holes. Other columns provide several examples of the numerical values of escape probabilities [Eq.~\eqref{AreaCapture}] for non-extremal black holes.}
  \label{table:numericalresults}
\end{table}

\subsubsection{A near-NHEK ZAMS}\label{sec:epnnk}
In the backside projected area-preserving plane \eqref{DefbackProjection}, the critical curve for a ZAMS at the IPS \eqref{InnerShell} in the near-NHEK region [Eq.~\eqref{nnkCritPsifull}] is obtained as
\begin{align}
\label{nnkIMCritRho}
	\td\rho=
	\begin{cases}
	\sqrt{2+\frac{2\sin\td\varphi}{\sqrt{3+\sin^2\td\varphi}}}\qquad
&-\arctan\frac{1}{2}
\leq\td\varphi\leq\arctan\frac{1}{2},	 \\
    \sqrt{\frac{16}{7-3\sin^2\td\varphi}}\qquad
&-\frac{\pi}{2}\leq\td\varphi\leq-\arctan\frac{1}{2},\quad
\arctan\frac{1}{2}\leq\td\varphi\leq\frac{\pi}{2},\\
     2\qquad
&\frac{\pi}{2}\leq\td\varphi\leq\frac{3\pi}{2}.
	\end{cases}
\end{align}
Then from Eq.~\eqref{DefEP} Eq.~\eqref{AreaCapture} we obtain
\be
\label{nnkIMEP}
\mathcal{P}_e(r_{\text{IPS}})=\frac{5}{12} -\frac{1}{\sqrt{7}} + \frac{2}{\sqrt{7}\pi}\arctan\frac{1}{\sqrt{7}}\approx12.57\% .
\ee
This is the leading-order escape probability in $\kappa$ \eqref{SmallPara} for a ZAMS at the IPS of a near-extremal black hole\footnote{A precisely extremal black hole has $\kappa=0$ (corresponding to $\k\tau=0$ in \eqref{nearextremality}), thus the IPS degenerates onto the horizon and the near-NHEK region disappears, the near horizon geometry is instead resolved by performing an infinite dilation on the horizon which results in the NHEK geometry (see Appendix~\ref{app:nhgeometry}) \cite{Kapec:2019hro}. \label{ftnote}}.
Note that the proper radial distance between the horizon and the IPS to the leading order in $\kappa$ is
$M\log\sqrt{3}$ \cite{Bardeen:1972fi}, which is a finite value and thus the ZAMS at IPS is sufficiently close to the horizon (compare to the extremely deep throat).
Therefore, this result confirms the conclusion in \cite{Ogasawara:2019mir} that the escape probability in a high spin black hole spacetime keeps a relatively large non-zero value as the ZAMS approaches to the horizon until it reaches just outside the horizon. Moreover, we find that the IPS \eqref{InnerShell} could be regarded as an exact bound (at the horizon scale) for a ZAMS at which the escape probability tends to a considerable finite value, $12.57\%$. We show this bound in Fig.~\ref{fig:throat} and Fig.~\ref{fig:EscapeProbability} and present a number of numerical results for various near-extremal spins in Table \ref{table:numericalresults} (we show also an exception of high spin, 0.7, for comparison).

It has been shown \cite{Gralla:2017ufe} that  the photons escaping from the NHEK and near-NHEK region of a high-spin black hole can all arrive at a vertical NHEKline on an observer's screen in the far region.
This means that the physical processes happened near a high-spin black hole at the horizon scale are in principle visible and may produce rich signals on the NHEKline.

\section{Energy to infinity}\label{sec:redshift}
\begin{figure}[tp]
  \centering
  \includegraphics[width=17cm]{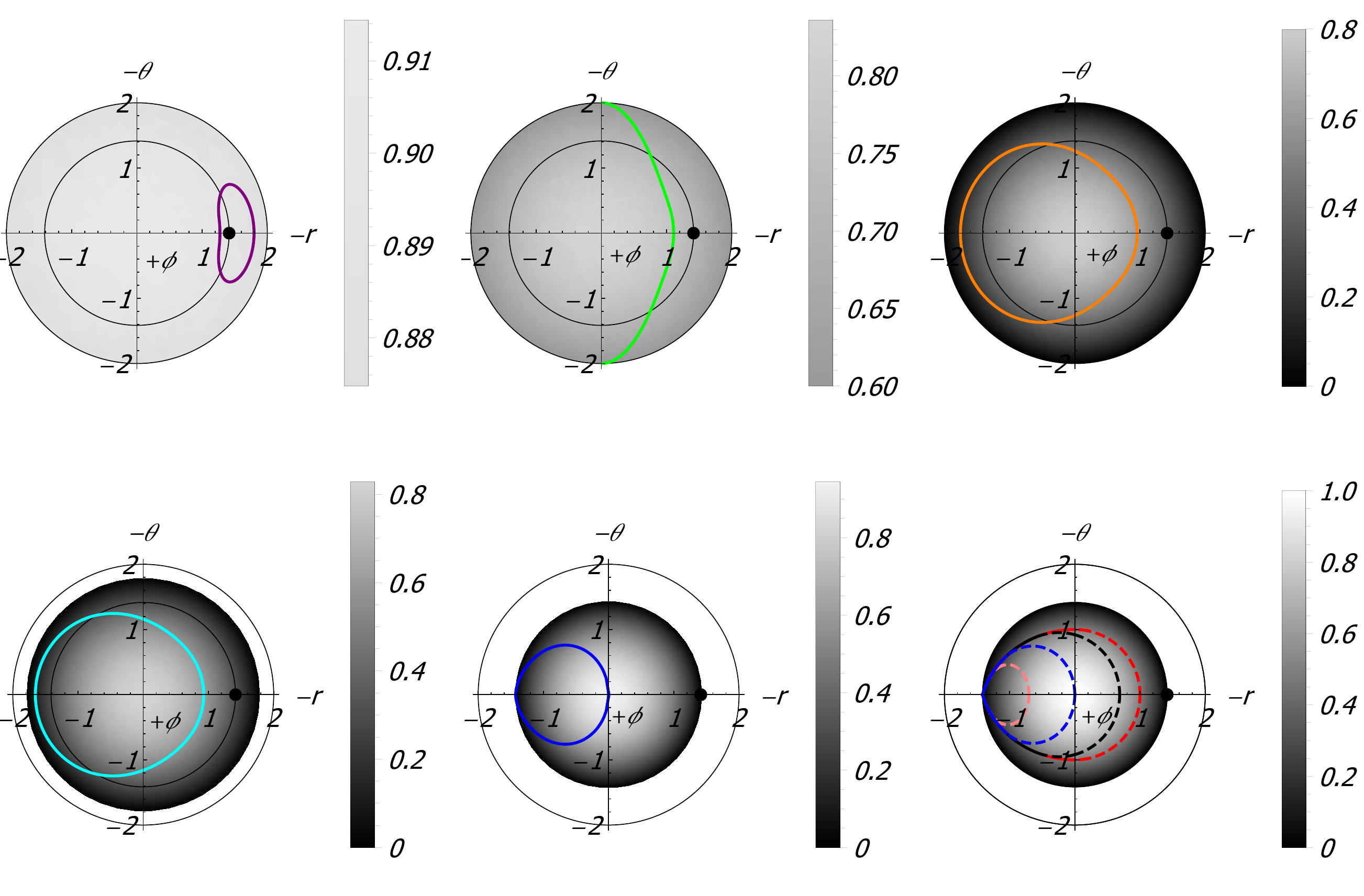}
  \caption{Redshift density plot in the frontside projected plane \eqref{DefbackProjection}: the center $\rho_f=0$ has $\Psi=0$, the middle circle $\rho_f=\sqrt{2}$ has $\Psi=\pi/2$ and the outer circle $\rho_f=2$ has $\Psi=\pi$. The first five plots are for $a=0.998M$ [Eqs.~\eqref{gofpsi} and \eqref{psi0}] while the last one is for $a\rightarrow M$ [Eq.~\eqref{gofpsink}], and the color-coding for the critical curves are the same as in Fig.~\ref{fig:EscapeCone}. In each plot, the black dot represents the ``direction to the black hole center"  and the photons in the same region (divided by the colored critical curve) with it are captured by the central black hole. Note that for a ZAMS in the ergoregion (second line), the photon emissions from the region $\Psi>\Psi_0$ \eqref{psi0} in the source sky have (formally) negative redshift factor and we show such regions with blank density in the plots.}
  \label{fig:RedshiftDensity}
\end{figure}

The observed energy of escaping photons at infinity is shifted from local emission energy. The redshift factor is defined by
\bea
\label{gofpsi}
g=\frac{E}{p^{(t)}}
=\frac{ r\sqrt{\Delta(r_s)}+2aM\cos\Psi}{\sqrt{r_s\xi_s}},
\eea
where the second equality is obtained by using Eq.~\eqref{eq:CosPsi}.
It is easy to see that the redshift factor $g$ decreases monotonically with $\Psi$ since
\be
\frac{\partial g(\Psi)}{\partial\Psi}=-\frac{2aM\sin\Psi }{\sqrt{r_s\xi_s}}<0.
\ee
For $g(\Psi_0)=0$ we have
\be
\label{psi0}
\Psi_{0}=\arccos\br{-\frac{r_s\sqrt{\Delta(r_s)}}{2aM}}.
\ee
Note that a physical angle $\Psi_0$ requires $\ab{\cos\Psi_0}\leq1$, which is equivalent to $r_s\leq2M$. In that case, there is a region in the source sky in which the redshift factor becomes (formally) negative for $\Psi>\Psi_0$ and thus the photon emissions from that region are not relevant to observations since the observed energy at infinity can never be negative.
The largest angle for the photon emissions with positive redshift factor is
then given by
\begin{align}
	\Psi_{\text{max}}=
	\begin{cases}
	\Psi_0\qquad
&r_s<2M,	 \\
		\pi\qquad
&r_s\geq2M .
	\end{cases}
\end{align}

Next, we examine the high-spin limits of the redshift factor and the maximum angle with non-negative redshift.
In the throat of an extremal Kerr black hole, we have
\be
\label{gogpsiek}
g=\frac{r^2_s-Mr_s+2M^2\cos\Psi}{\sqrt{r_s(2M^3+M^2r_s+r_s^3)}},\qquad
\Psi_{\text{max}}=\arccos\br{-\frac{(r_s-M)r_s}{2M^2}}.
\ee
In the NHEK and near-NHEK limit, we have
\be
\label{gofpsink}
g=\cos\Psi,\qquad
\Psi_{\text{max}}=\frac{\pi}{2}.
\ee

In contrast to the case for circular geodesic motion \cite{Igata:2019hkz,Gates:2020sdh,Gates:2020els}, no photon emitted from a ZAMS could have a net blueshift when it reaches to asymptotic infinity. The reason for this is that the ZAMS may be treated as a relatively ``static" source which has no transverse Doppler boost, while for the source in a circular motion the proper motion in $+\phi$ direction of a circular source is at relativistic speed and the Doppler boost overcomes the gravitational redshift. We show the redshift density plot in the frontside projected plane of the source sky in Fig.~\ref{fig:RedshiftDensity}.

\section{Summary and remarks}\label{sec:discussion}
In this work, we  considered monochromatic and isotropic photon emissions from ZAMSs near a Kerr black hole. We  computed the escape probabilities and the energy shifts of escaping photons and expressed them as the functions [Eqs.~\eqref{DefEP}, \eqref{AreaCapture} and \eqref{gofpsi}] of the source radius $r_s$ and the spin of the black hole $a$.
The escape probabilities have been summarized in Fig.~\ref{fig:EscapeProbability} and in Table \ref{table:numericalresults}.
We can see that the escape probability decreases as $r_s$ approaches to the horizon for each given value of spin $a$, and it decreases as $a$ increases at the IPS or at the ISCO, respectively.
These results agree with those found in \cite{Ogasawara:2019mir}, but we have obtained them in a different manner introduced in \cite{Gates:2020sdh}.
Furthermore, we used the high-spin perturbation method to study the (near-)extremal case and computed the escape probabilities and redshift factors to the leading order in the deviation from extremality, paying special attention to the NHEK ZAMSs and near-NHEK ZAMSs. The escape probability for high-spin cases at the Thorne limit $a=0.998M$ and at the extremal limit $a\rightarrow M$ are summarized in Fig.~\ref{fig:throat} and Fig.~\ref{fig:EscapeProbability}. As $a\rightarrow M$, the escape probability tends to $29.17\%$ at ISCO (with divergent proper distance to the horizon, $\sim\log\kappa^{-1}$) and tends to $12.57\%$ at IPS (with finite proper distance to the horizon, $M\log\sqrt{3}$). This clarifies the inner bound \cite{Ogasawara:2019mir} in the near-horizon region with considerable observability which is useful for exploring the near-horizon physics of a high-spin black hole.
In addition, we provided several redshift density plots  in Fig.~\ref{fig:RedshiftDensity}, showing that the gravitational redshift dominates over the Doppler boost for the photon emission from ZAMSs.

The  escape probability and the redshift of the photon from a near-NHEK ZAMS show that the deepest region in a near-horizon throat is visible to distant observers in principle, even though
the observability of a ZAMS is suppressed by the strong gravity effect due to the zero angular momentum of its motion.
Observing such sources may give us insights to understand the super-Penrose process occurred in the ergoregion. On the other hand, for the sources with other motions, the observability are supposed to be enhanced by the Doppler effect due to their relativistic speeds.
The possible orbits of light sources with relativistic motions include the plunging orbits inside the ISCO \cite{Igata:2021njn}, the unstable circular orbits between IBCO and ISCO, and elliptical orbits  inside an elliptical disk \cite{Liu:2020ljq}. A careful study for these kinds of sources in the NHEK and near-NHEK regions is deferred to future works.

In this paper, the observability of a ZAMS in the most general case has been evaluated only numerically. On the other hand, for the high-spin case, we have analytically performed a high-spin expansion \eqref{SmallPara} but only to the leading order in $\kappa$. In order to improve the results for a high-spin black hole with finite derivation from extremality, the corrections of
higher orders must be incorporated, this requires one to proceed the high-spin expansions to the subleading orders \cite{Li:2020val,Gates:2020sdh}.

\section*{Acknowledgments}
We thank Peng-Cheng Li and Zezhou Hu for their helpful discussions and comments on the manuscript.
The work is in part supported by NSFC Grant  No. 11735001. MG is also funded by China Postdoctoral Science Foundation Grant No. 2019M660278 and 2020T130020.

\appendix
\section{Near horizon geometry of (near-)extreme Kerr}\label{app:nhgeometry}
In this appendix we give a brief review on the NHEK and near-NHEK geometry which is sufficient for understanding the current paper. For a comprehensive review readers may refer to Ref.~\cite{Kapec:2019hro}.

\subsection{Extreme Kerr and NHEK}\label{app:NHEK}
Setting $a=M$ in Eq.~\eqref{eq:Metric1} gives the Extreme Kerr metric which has some peculiar features \cite{Bardeen:1972fi,Bardeen:1999px,Kapec:2019hro}. To investigate this extremal limit, it is helpful to consider
\be
a=M\sqrt{1-\kappa^2},\qquad 0\leq\kappa\ll1,
\ee
and take the limit $\kappa\rightarrow0$. Several puzzles arise immediately when looking carefully at the ISCO:
on one hand, we see that the ISCO coincides with the horizon in the Boyer-Lindquist coordinates
\be
\lim_{\kappa\rightarrow0}r_{\text{ISCO}}=M=\lim_{\kappa\rightarrow0}r_+;
\ee
on the other hand, the proper distance between the ISCO and the horizon as measured on a Boyer-Lindquist time-slice diverges
\be
\lim_{\kappa\rightarrow0}ds(r_+,r_{\text{ISCO}})\sim M\ab{\log\kappa}=\infty.
\ee
Therefore, the Extreme Kerr misrepresents the near-horizon geometry of an extremal Kerr black hole which appears as an infinitely deep throat.

In order to resolve this infinitely deep near-horizon throat we need to work with the Bardeen-Horowitz coordinates \cite{Bardeen:1999px}
\be
\label{BHcoordinate}
R=\frac{r-r_+}{\ve\, r_+},\qquad
T=\ve\, \Omega_H t,\qquad
\Phi=\phi-\Omega_H t,
\ee
where
\be
\label{OmegaH}
\Omega_H=\frac{a}{2Mr_+}
\ee
denotes the angular velocity of the black hole horizon,
and take the limit $\ve\rightarrow 0$.
The resulting geometry is the NHEK metric \cite{Bardeen:1999px}
\be
\label{NHEKmetric}
ds^2=2M^2\Gamma\br{-R^2 dT^2+\frac{dR^2}{R^2}+d\theta^2+\Lambda^2(d\Phi+RdT)^2},
\ee
where
\be
\Gamma(\theta)=(1+\cos^2\theta)/2,\hspace{3ex} \Lambda(\theta)=2\sin\theta/(1+\cos^2\theta).
\ee
 Note that the NHEK metric is $\ve-$independent and further coordinate transformation, $(R, T)\rightarrow (\ve R, T/\ve)$, leave the metric invariant. This means that the NHEK geometric enjoys a dilation symmetry due to its infinite depth. The NHEK coordinate $R=\infty$ corresponds to the entrance to the throat where the NHEK region glues on to the asymptotically flat Kerr geometry.

\subsection{Near-horizon limits for near-Extreme Kerr}\label{app:nearNHEK}
For near-extremal Kerr black hole we introduce \cite{Kapec:2019hro}
\be
\label{nearextremality}
a=M\sqrt{1-(\kappa \tau)^2},
\qquad
0<\kappa\ll1
\ee
to describe the deviation from extremality\footnote{Here $\tau$ is a finite value corresponding to a near-horizon temperature, which is related to the Hawking temperature $T_H$ by \cite{Bredberg:2009pv}
\be
\tau\equiv\frac{4\pi M T_H}{\ve}=\frac{1}{\ve}\frac{r_+-r_-}{2r_+}.\nn
\ee
We have chosen $\tau=1$ in the main text for convenience.}, under which the event horizon is given by
\be
\label{nearNHEKhorizon}
r_+=M(1+\kappa\tau).
\ee
Unlike the near horizon region for a precisely extremal black hole, the near horizon region for a near-extremal black hole is not unique. To make it clear, it is helpful to rewrite the near horizon scaling $\ve$ [see Eq.~\eqref{BHcoordinate}] as \cite{Hadar:2014dpa,Gralla:2015rpa}
\be
\label{deltaofkappa}
\ve=\kappa^p,\qquad 0<p\leq1.
\ee
Then we consider two different near horizon radii $r_{i}=r_+(1+\kappa^{p_i} R_{i}),\, (i=1,2)$, with $p_1\leq p_2$. It turns out that the proper radial distance between $r_1$ and $r_2$ is finite only for $p_1=p_2$ and it becomes divergent for $p_1\neq p_2$ (scale as $\ab{\log\kappa^{p_2-p_1}}$) \cite{Kapec:2019hro}.

We will work again with the Bardeen-Horowitz coordinates \eqref{BHcoordinate} and \eqref{OmegaH} when zooming into the near-horizon region of a near-extremal black hole with \eqref{nearextremality}, and take the near horizon scaling as \eqref{deltaofkappa}.
Taking the limit $\kappa\rightarrow 0$ for $p=1$, which corresponds to zooming in the near-horizon region at the same rate as black hole spinning towards extremality,  gives the so-called near-NHEK geometry \cite{Bredberg:2009pv,Gralla:2015rpa}
\be
\label{nearNHEKmetric}
ds^2=2M^2\Gamma\br{-(R(R+2\tau)) dT^2+\frac{dR^2}{R(R+2\tau)}+d\theta^2+\Lambda^2(d\Phi+(R+\tau)dT)^2}.
\ee
While taking the limit $\kappa\rightarrow 0$ for $0<p<1$, which amounts to zooming into the near-horizon region at a rate slower than black hole spinning towards extremality, gives the same NHEK metric\footnote{Taking limit $\kappa\rightarrow0$ for $``p=0"$ gives again the Extreme Kerr metric which describes the region outside the near-horizon throat.} \eqref{NHEKmetric}. However, each specific value of $p$ corresponds to a distinct NHEK band, which separates with another considerably. From the perspective of near-extremality, the precisely extremal black hole has $\tau=0$ and thus the near-NHEK band, together with all the NHEK bands, reduces into one unique NHEK\footnote{In this case, $\k$ enters into the near-horizon scaling from \eqref{deltaofkappa} but has nothing to do with near-extremality.} \cite{Kapec:2019hro}.

Note that the event horizon \eqref{nearNHEKhorizon} and the IPS \eqref{ExtremeInner} scale into the near-NHEK band, and the ISCO scales into the $p=2/3$ NHEK band \cite{Gralla:2015rpa}. Examples of interesting proper radial distances among these orbits and the ergosurface \eqref{eq:ErgoSurface} are \cite{Kapec:2019hro}
\bea
\label{properdistance}
ds(r_+,r_{\text{IPS}})&=&M\log\sqrt{3}+\mathcal{O}(\kappa),\\
\label{rhrisco}
ds(r_+,r_{\text{ISCO}})&=&\frac{M}{3}\log\frac{2^4}{\kappa\tau}+\mathcal{O}
(\kappa^{2/3}),\\
\label{riscor0}
ds(r_{\text{ISCO}},r_{0+})&=&M\br{1+\frac{2}{3}\log\frac{1}{\sqrt{2}\kappa\tau}}
+\mathcal{O}(\kappa^{2/3}).
\eea

\bibliographystyle{utphys}
\bibliography{noteZAMS}



\end{document}